\documentclass[pra,twocolumn,superscriptaddress,floatfix,longbibliography]{revtex4-1}%nofootinbib,floatfix

\newcommand{\ket}[1]{|{#1}\rangle}
\newcommand{\bra}[1]{\langle{#1}|}

\newcommand{\Rho}{\mathrm{P}}

\usepackage{amsfonts}
\usepackage{amsmath}
\usepackage{graphicx}
\usepackage{amssymb}
\usepackage{color}
\usepackage{subfigure}
\usepackage[perpage]{footmisc}
\usepackage{mathtools}
\usepackage{lipsum}
\usepackage{lineno}
\usepackage{comment}
\usepackage{dsfont}
\begin{document}
\title{Particle creation in the spin modes of a dynamically oscillating two-component Bose-Einstein condensate}
%\title{Dynamical amplification of the zero-point fluctuations in a two-component condensates}
%\title{Vacuum amplification in a non-stationary two-component condensates}

\author{Salvatore Butera}
\affiliation{School of Physics and Astronomy, University of Glasgow, Glasgow G12 8QQ, UK}
\author{Iacopo Carusotto}
\affiliation{INO-CNR BEC Center and Dipartimento di Fisica, Universit\`a di Trento, I-38123 Povo, Italy}
\begin{abstract}
We investigate the parametric amplification of the zero-point fluctuations in the spin modes of a two-component Bose-Einstein condensate,  triggered by the dynamical evolution of the condensate density. We first make use of a Thomas-Fermi approximation to develop a tractable theoretical model of the quantum dynamics of the Bogoliubov excitations in a harmonically trapped condensate with a time-dependent trapping frequency. The predictions of this model are then compared to an ab-initio numerical study of the correlation functions of density and spin fluctuations for general spatially inhomogeneous configurations. Results are shown for the two cases of expanding and oscillating condensates: while the quantum excitation of spin modes remains weak and relatively featureless in the case of an expanding condensate, clear and experimentally promising signatures of particle creation are anticipated for the oscillating case under suitable resonance conditions between the density and the spin modes.
\end{abstract}
%\pacs{}
\maketitle

\section{Introduction}

It is well known that the zero point fluctuations of a quantum field can be excited into observable radiation in the case of a time-dependent or, more generally, curved spacetime \cite{birrell1984quantum}. Examples of such phenomena for non-stationary backgrounds are the cosmological particle creation \cite{Parker-PartCr-I,Parker-PartCr-II} and the dynamical Casimir effect (DCE) \cite{Moore-DCE-1970,Fulling_Davies-DCE,Dodonov-DCE-Rev}. In the former case the non-stationarity is in the metric, that is a bulk property of the spacetime while, in the latter case, the time-dependence is in a boundary condition imposed to the field. To the same family of phenomena belongs also the Hawking radiation emanating from black holes \cite{Hawking1974,Hawking1975}. In this case however, the emission originates from the presence of an event horizon and is thus linked to a modification of the causal structure of spacetime.

The detection of tiny quantum effects in a cosmological context  is extremely challenging with state-of-the-art technologies. So far, the only (indirect) evidence is in the anisotropy of the cosmic microwave background \cite{Hu-CMB-1996} that, according to the theory of Cosmological inflation \cite{Bassett-RMP-Inflation}, is believed to be a signature of primordial vacuum fluctuations in the early Universe. These difficulties have pushed for the quest of \emph{analog} systems \cite{Barcelo-2011,faccio-book}, where the microscopic physics is different from gravity, but the same kinematic effects of quantum field theory on  time-dependent or curved backgrounds can be implemented and tested in a lab. A surge of proposal have flourished over the past couple of decades using a multitude of analog systems, including Bose-Einstein condensates (BEC) of ultracold atoms~\cite{Garay2000,Carusotto2008,Recati2009,finazzi2014,Hu-Cosm-BEC-2003,Fedichev2003,Fedichev2004,Uhlmann2005,Jain-Cosm-BEC-2007,Prain-Cosm-BEC-2010}, ions \cite{Schutzold-Cosm-Ions-2007,Schtzhold-Cosm-Ions-2018,Schtzhold-Cosm-Ions-2019}, quantum fluids of light \cite{Gerace2012},  and superconducting circuits \cite{Schutzhold2005,Nation-PRL-2009,Schtzhold-Cosm-Circuits-2019} to name a few. Building on this theoretical effort, pioneering experimental works claimed the detection of spontaneous Hawking emission originating from a sonic black hole \cite{steinhauer2016,steinhauer2019,steinhauer2021} or from effective horizons in a nonlinear medium \cite{Faccio-PRL-2010}, or of its classical, stimulated counterpart in surface waves on water \cite{Rousseaux-PRL-2016,Rousseaux-PRL-2020}. Experimental studies of superradiant scattering in rotational geometries~\cite{Silke-SuperRad-Nature} and of particle creation in analogs of an expanding Universe~\cite{Hung1213,Eckel-Cosm-BEC-2018,steinhauer2021-Light} have also been reported.

In all these works, the non-stationary effective spacetime is simulated by externally modulating certain physical properties of the system at hand, such as the refractive index in a optical medium or the scattering length of the two-body collisional interaction in an atomic BEC. In other words, in such proposals the time-dependence driving the parametric amplification of the vacuum fluctuations is not provided by a dynamical degree-of-freedom of the system, but is rather imposed by the external action of the experimentalist. While this approach is sufficient to study kinematic effects of quantum field theory on a curved spacetime, it can not be used to go beyond and address those dynamical and back-reaction features that are more and more attracting the interest of the community~\cite{BeiLok-book}.

In this work we consider a conceptually different configuration in which the vacuum fluctuations get amplified by the dynamical evolution of the system itself. By either switching off, inverting, or just suddenly perturbing the frequency of the harmonic trapping, different behaviours can be generated in the condensate such as a linear or exponential expansion, or periodic oscillations. In the cosmological analogy, these regimes simulate an expanding or a more complex cyclic universe ~\cite{Jain-Cosm-BEC-2007,Turok_Cyclic} or the last preheating stage of inflation~\cite{PreHeating-Rev}.
Capitalizing on previous works~\cite{Fischer2004,Visser2005,liberati2006analogue}, we focus our attention on the most promising case of an analog model based on a two-component BEC. The spinorial nature of the BEC gives rise to two independent branches of collective excitations which, in the simplest spin symmetric case, have purely density or spin characters~\cite{Abad2013}. Going beyond our classical study of black hole lasing dynamics of spin waves in~\cite{Butera2017}, here we investigate quantum particle creation processes into the spin excitation branch that are driven by the dynamical evolution of the overall condensate density modes. Within this framework, the density excitations play the role of the non-stationary background (namely the spacetime in the gravitational case), while the spin excitation modes encode the quantum field. A key advantage of this configuration is that the speed of standard (density) sound can be much faster than the one of spin-sound, so that one can take advantage of the faster characteristic time scale of the density oscillations to enhance the particle production into the spin modes. Further experimental advantages of spinor condensates are offered by the possibility of simultaneous imaging both the density and the spin profiles in real time~\cite{farolfi2020quantum}.

The work is organized as follows: In Sec.~\ref{Sec:TwoComponentBEC}, we develop a theoretical model under the simplifying assumption of a spatially homogeneous system. We show that this model is able to predict the amplification of the vacuum fluctuations in the spin excitation modes as a result of the time-dependent overall density. We derive the effective action for the fluctuations and show that, in the case of a expanding background, both the phase and the density experience an effective damping. In Sec.~\ref{Sec:Homog} we make use of this simplified model to simulate the dynamics of the quantum fluctuations in a harmonically trapped condensate within the Thomas-Fermi limit in different cases of a linearly or exponentially expanding condensate and of an oscillating one. In Sec.~\ref{Sec:InHom}, we present an ab-initio numerical study of the dynamics of the trapped system, focusing on the time evolution of the two-body correlations in the density and the sectors. While the signal of the parametric amplification of the vacuum fluctuations remains weak in an expanding condensate, strong signatures are instead found in the case of an oscillating condensate. Our final considerations and our perspectives for future work are finally summarized in Sec.~\ref{Sec:Conclusions}.

\begin{figure*}[!htbp]
\centering
\includegraphics[width=0.9\textwidth]{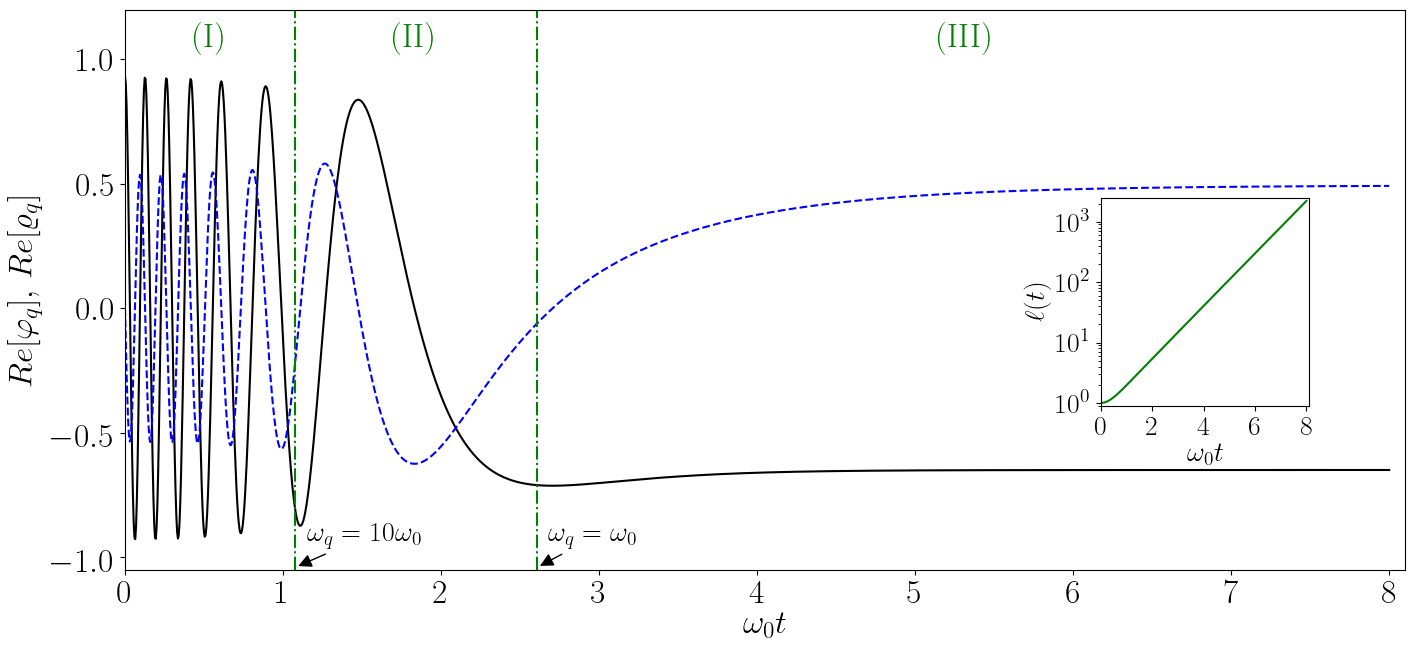}
\caption{Time evolution of the (spin) density (solid black line) and (relative) phase (dashed blue line) components of a spin mode of initial angular frequency $\omega_s/\omega_0 = 50$, in a one-dimensional condensate that is expanding after reverting the sign of the harmonic potential at $t=0$. In the inset we show the dynamics of the expansion parameter $\ell(t)$: After the initial transient regime, whose duration is of the order $\sim1/\omega_0$, the expansion is exponential, and characterized by the Hubble parameter $H(t)=\dot{\ell}(t)/\ell(t) = \omega_0$. The three stages of the evolution of the modes discussed in the main text, driving the system from the under- to the over-damped regime, are clearly visible.}
\label{Fig:2}
\end{figure*}

\section{Theory of a non-stationary two-component condensate\label{Sec:TwoComponentBEC}}
\subsection{Lagrangian and energy functional}
We consider a weakly interacting $D-$dimensional Bose gas composed by two atomic species $a$ and $b$ or two different internal atomic states of the same atom~\cite{Stringari_BEC}. The atoms are assumed to have the same mass $m$ in the two states and to be subject to the same external harmonic potential $V(\mathbf{x},t) = (m/2)\sum_{\substack{i=1}}^D\omega_i(t)x_i^2$, with generic time-dependent trapping frequencies $\omega_i(t)$ in the three directions $i=\{x,y,z\}$. We indicate by $g_{jj'}$ $(j,j'=a,b)$ the interaction constants for the different collisional channels. 

The action $S$ describing the dynamics of the system is expressed as the space-time integral of the Lagrangian density $\mathcal{L}$, that is
\begin{equation}
	S=\int{dt\,d\mathbf{x}\, \mathcal{L}\,(\hat{\Psi}_j,\hat{\Psi}_j^\dag,\partial_\alpha\hat{\Psi}_j,\partial_\alpha\hat{\Psi}_j^\dag)}.
\label{Eq:Action}
\end{equation}
Here, $d\mathbf{x}\equiv\prod_{i=1}^D dx_i$ is the differential volume element, $\hat{\Psi}_j(\mathbf{x},t)$ $(j=a,b)$ are the field operators relative to the two components of the system, and we collectively indicated the space and time derivatives by using the $D+1$ notation $\partial_\alpha\hat{\Psi}_j$ ($\alpha=0,1,...,D$, the time coordinate corresponding to $\alpha=0$). The Lagrangian density can be written in terms of the density $\hat{\Rho}_j(\mathbf{x},t)$ and phase $\hat{\Phi}_j(\mathbf{x},t)$ operators, which are defined according to the Madelung representation of the fields $\hat{\Psi}_j=\sqrt{\hat{\Rho}_j}e^{i\hat{\Phi}_j}$. By using this notation, the Lagrangian takes the explicit form:
\begin{multline}
	\mathcal{L}(\hat{\Rho}_j,\hat{\Phi}_j,\partial_\alpha\hat{\Rho}_j,\partial_\alpha\hat{\Phi}_j)=\\ \sum_{j=a,b}{\left(\hbar\hat{\Rho}_j\frac{\partial\hat{\Phi}_j}{\partial t}\right)}+\epsilon(\hat{\Rho}_j,\hat{\Phi}_j,\partial_\alpha\hat{\Rho}_j,\partial_\alpha\hat{\Phi}_j),
\label{Eq:LagrDens}
\end{multline}
where
\begin{multline}
	\epsilon(\hat{\Rho}_j,\hat{\Phi}_j,\partial_\alpha\hat{\Rho}_j,\partial_\alpha\hat{\Phi}_j)=\sum_{\substack{j=a,b}}\left[\frac{\hbar^2}{8m\hat{\Rho}_j}\left(\boldsymbol{\nabla}_\mathbf{x}\hat{\Rho}_j\right)^2\right.\\
	+\left.\frac{\hbar^2}{2m}\hat{\Rho}_j(\boldsymbol{\nabla}_\mathbf{x}\hat{\Phi}_j)^2+V_j(\mathbf{x},t)\hat{\Rho}_j+\frac{g_{jj}}{2}\hat{\Rho}_j^2+\frac{g_{jj'}}{2}\hat{\Rho}_j\hat{\Rho}_{j'}\right]
\label{Eq:EnFunc}
\end{multline}
is the energy density of the system. In Eq.~\eqref{Eq:EnFunc}, $\boldsymbol{\nabla}_\mathbf{x}$ is the standard \emph{nabla} operator, that is the vector-valued differential operator whose components are the derivative respect to each spatial coordinate. From now on, we indicate with the primed index $j'$ the component of the system other than $j$.

\subsection{Co-moving coordinates\label{Subsec:Coord}}
We describe the evolution of the system by working with the so-called \emph{co-moving} coordinates $y_i\equiv x_i/\ell_i(t)$ $(i=1,...,D)$ \cite{Castin-ComCoord-1996,Shlyapnikov-ComCoord-1996,Shlyapnikov-ComCoord-1997}, in which the expansion parameters $\ell_i(t)$ account for the size variation of the system, and defined the scaling volume $\mathcal{V}(t)\equiv\prod_{i=1}^D \ell_i(t)$. An explicit evolution law for $\ell_i(t)$ will be given in Eq.~\eqref{Eq:B_Evol}.
For notational convenience, we also introduce the rescaled operators $\hat{\phi}_j$, $\hat{\rho}_j$ and $\hat{\psi}_j$, defined according to the following transformations:
\begin{subequations}
\begin{align}
	\hat{\Phi}_j(\mathbf{x},t)&=%\rightarrow
	\sum_{\substack{i=1}}^D\frac{\dot{\ell}_i(t)}{2\ell_i(t)} \frac{\hbar}{m} x_i^2+\hat{\phi}_j(\mathbf{y},t),\label{Eq:ScaledPhi}\\
	\hat{\Rho}_j(\mathbf{x},t)&=%\rightarrow
	\frac{\hat{\rho}_j(\mathbf{y},t)}{\mathcal{V}(t)},\label{Eq:ScaledRho}\\
	\hat{\Psi}_j(\mathbf{x},t)&=%\rightarrow
	\frac{1}{\sqrt{\mathcal{V}(t)}}\exp\left[\sum_{\substack{i=1}}^D\frac{\dot{\ell}_i(t)}{2\ell_i(t)} \frac{i\hbar}{m} x_i^2\right]\hat{\psi}_j(\mathbf{x},t).\label{Eq:ScaledPsi}
\end{align}
\end{subequations}
The first term in Eq.~\eqref{Eq:ScaledPhi} accounts for the phase induced by the overall motion of the system while, in Eq.~\eqref{Eq:ScaledRho}, $\hat{\rho}_j$ is the scaled density profile. According to the definitions in Eqs.~\eqref{Eq:ScaledPhi}-\eqref{Eq:ScaledPsi}, the scaled field operators are related via the usual Madelung relation $\hat{\psi}_j=\sqrt{\hat{\rho}_j}\exp(i\hat{\phi}_j)$.

The action $S$ can be written in terms of these scaled quantities as \cite{Fedichev2004,Castin-ComCoord-1996}
\begin{multline}
	S=\int{dt\,d\mathbf{y}\sum_{\substack{j=a,b}}}\left\{\hbar\hat{\rho}_j\frac{\partial \hat{\phi}_j}{\partial t}+\right. \\ \left.+\sum_{\substack{i=1}}^D\left[\frac{1}{2}m(\omega_{i,0}y_i)^2\frac{\hat{\rho}_j}{\mathcal{V}(t)}+\frac{\hbar^2}{8m\hat{\rho}_j}\left(\frac{1}{\ell_i(t)}\frac{\partial\hat{\rho}_j}{\partial y_i}\right)^2\right.\right.+ \\
	{\left.\left.+\frac{\hbar^2}{2m}\left(\frac{1}{\ell_i(t)}\frac{\partial\hat{\phi}_j}{\partial y_i}\right)^2\hat{\rho}_j\right]+\frac{g_{jj}}{2}\frac{\hat{\rho}_j^2}{\mathcal{V}(t)}+\frac{g_{jj'}}{2}\frac{\hat{\rho}_j\hat{\rho}_{j'}}{\mathcal{V}(t)}\right\}},
\label{Eq:ActionScaled}
\end{multline}
where we indicated the initial value of the trapping frequency $\omega_{i,0} = \omega_i(0)$.

Within the usual Thomas-Fermi interaction, valid if the interaction energy is much larger than the harmonic trap frequency~\cite{Stringari_BEC}, the dynamics of the scale parameters $\ell_i(t)$ is governed by the equation \cite{Castin-ComCoord-1996}
\begin{equation}
	\ddot{\ell}_i(t)+\omega_{i}^2(t)\ell_i(t)=\frac{\omega_{i,0}^2}{\ell_i(t)\mathcal{V}(t)},
\label{Eq:B_Evol}
\end{equation}
which, for a system initially at equilibrium, has to be solved with the initial conditions $\ell_i(0)=1$, $\dot{\ell}_i(0)=0$. We indicate time derivatives with over dots.

\subsection{Bogoliubov theory\label{Subsec:Fluct}}
\noindent We follow the Bogoliubov prescription, and split the field operators into their mean-field (classical) $\phi_{j,0},\,\rho_{j,0}$ and quantum fluctuating $\delta\hat{\phi}_j,\,\delta\hat{\rho}_j$ components as
\begin{subequations}
\begin{align}
\hat{\phi}_j&=\phi_{j,0}+\delta\hat{\phi}_j,\label{Eq:BogoAnsatz_a_Com}\\
\hat{\rho}_j&=\rho_{j,0}+\delta\hat{\rho}_j.\label{Eq:BogoAnsatz_b_Com}
\end{align}
\end{subequations}
Accordingly, we expand the action in Eq.~\eqref{Eq:ActionScaled} up to second order in $\delta\hat{\phi}_j,\delta\hat{\rho}_j$, in order to capture the free dynamics of the quantum fluctuations. 

\subsubsection{Condensate evolution}
\noindent The zero-th order term of the action has the same structure as Eq.~\eqref{Eq:ActionScaled}, with the operators replaced by their mean-field components. This provides the following Euler-Lagrange equations for the phase and the density:
\begin{multline}
    -\hbar\frac{\partial\phi_{j,0}}{\partial t} = \frac{m}{2}\sum_{\substack{i=1}}^D{\left(\omega_{i,0}^2\frac{y_i^2}{\mathcal{V}(t)}\right)}+\\
    +\sum_{\substack{i=1}}^D\left[\frac{\hbar^2}{2m}\left(\frac{1}{\ell_i(t)}\frac{\partial\phi_{j,0}}{\partial y_i}\right)^2-\frac{\hbar^2}{4m\ell_i^2(t)}\frac{\partial}{\partial y_i}\left(\frac{1}{\rho_{j,0}}\frac{\partial\rho_{j,0}}{\partial y_i}\right)\right]+\\
    +g_{jj}\frac{\rho_{j,0}}{\mathcal{V}(t)}+g_{jj'}\frac{\rho_{j',0}}{\mathcal{V}(t)},
\end{multline}
\begin{equation}
     \frac{\partial\rho_{j,0}}{\partial t} + \frac{\hbar}{m} \sum_{\substack{i=1}}^D\left[\frac{\partial}{\partial y_i}\left(\frac{1}{\ell_i(t)}\frac{\partial\phi_{j,0}}{\partial t}\rho_{j,0}\right)\right]= 0.
\end{equation}
In the Thomas-Fermi (TF) limit in which the spatial variations of the density can be neglected~\cite{Stringari_BEC}, these are solved by taking a time-independent value $\rho_{j,0} $ of the scaled  density such that
\begin{equation}
	\mu=\frac{m}{2}\sum_{\substack{i=1}}^D\omega_{i,0}^2 y_i^2+g_{jj}\rho_{j,0}+g_{jj'}\rho_{j',0}.
	\label{Eq:EqDens}
\end{equation}
and a simple evolution of the spatially-uniform scaled phase in the form
$\phi_{j,0}(t) = -{\mu\tau(t)}/{\hbar}$ \cite{Castin-ComCoord-1996,Fedichev2004}, where $\mu$ is the initial chemical potential of the system and $\tau$ is a co-moving time defined according to the differential relation $d\tau=dt/\mathcal{V}$.

Throughout this paper, we focus on  the case of a symmetric system, for which the contact interaction strengths between particles in states $a$ and $b$ are the same $g_{aa} = g_{bb} \equiv g$. In this assumption, the mean-field ground state is symmetric or polarized, depending on whether %$g\leq g_{ab}$ or $g> g_{ab}$.
or $g> g_{ab}$ or $g<g_{ab}$.
In the former case the density for the two components is the same, and equal to $\rho_{a,0}=\rho_{b,0}=\rho_{0}/2$, with
\begin{equation}
	\rho_0 = \left(\frac{2\mu-m\sum_{i=1}^D\omega_{i,0}^2 y_i^2}{g+g_{ab}}\right).
\label{EQ:TF}
\end{equation}
In this symmetric configuration, the elementary excitations of the system decouple into the two independent \emph{spin} and \emph{density} branches \cite{Abad2013}.

\subsubsection{Collective excitations}

The equations governing the dynamics of the collective excitations on top of the condensate are obtained from the second order term of the action in Eq.~\eqref{Eq:ActionScaled} in the fluctuation operators. This has the form:

\begin{widetext}
\begin{multline}
	S^{\rm (2)}=\int{dt\,d\mathbf{y}\,\sum_{\substack{j=a,b}}\left\{ \hbar\delta\hat{\rho}_j\frac{\partial\delta\hat{\phi}_j}{\partial t}
	+\sum_{\substack{i=1}}^D\left[\frac{\hbar^2}{4m\rho_0}\left(\frac{1}{\ell_i(t)}\frac{\partial\delta\hat{\rho}_j}{\partial y_i}\right)^2
	-\frac{\hbar^2}{2m\rho_0^2}\left(\frac{1}{\ell_i(t)}\frac{\partial\rho_0}{\partial y_i}\right)\left(\frac{1}{\ell_i(t)}\frac{\partial\delta\hat{\rho}_j}{\partial y_i}\right)\delta\hat{\rho}_j \right.\right.} \\
	{\left.\left.+\frac{\hbar^2}{4m}\rho_0\left(\frac{1}{\ell_i(t)}\frac{\partial\delta\hat{\phi}_j}{\partial y_i}\right)^2\right]+\frac{g}{2\mathcal{V}(t)}\delta\hat{\rho}_j^2+\frac{g_{ab}}{2\mathcal{V}(t)}\delta\hat{\rho}_j\delta\hat{\rho}_{j'}\right\}}.
\label{Eq:ActionScaled_2nd}
\end{multline}
\end{widetext}

For simplicity, in this section we restrict our attention to the central region of the condensate, where the density can be approximated as homogeneous ($\partial\rho_0/\partial y_i\approx 0$). Under this assumption the action in Eq.~\eqref{Eq:ActionScaled_2nd} reduces to the form
\begin{multline}
	S^{\rm (2)}=\int{dt\, d\mathbf{y}\sum_{\substack{j=a,b}}\left\{
	\hbar\delta\hat{\rho}_j\frac{\partial\delta\hat{\phi}_j}{\partial t}\right.}\\
	+{\left.\sum_{\substack{i=1}}^D\left[\frac{\hbar^2}{4m\rho_0}\left(\frac{1}{\ell_i(t)}\frac{\partial\delta\hat{\rho}_j}{\partial y_i}\right)^2+\frac{\hbar^2}{4m}\rho_0\left(\frac{1}{\ell_i(t)}\frac{\partial\delta\hat{\phi}_j}{\partial y_i}\right)^2\right]\right.}\\
	{\left.+\frac{g}{2\mathcal{V}(t)}\delta\hat{\rho}_j^2+\frac{g_{ab}}{2\mathcal{V}(t)}\delta\hat{\rho}_j\delta\hat{\rho}_{j'}\right\}}.
\label{Eq:ActionScaled_2nd_Hom}
\end{multline}
where the fluctuations in the $a$ and $b$ components are coupled by the cross-species collisional interaction described by the last term in Eq.~\eqref{Eq:ActionScaled_2nd_Hom}. 

A further simplified form is obtained by working in the density and spin basis, $\delta\hat{\sigma}_d = (\delta\hat{\sigma}_a+\delta\hat{\sigma}_b)/2$ and $\delta\hat{\sigma}_s = (\delta\hat{\sigma}_a-\delta\hat{\sigma}_b)/2$ for both the density and the phase $\sigma = \rho,\,\phi$. While the $\rho_d$ and $\phi_d$ have the usual meaning of the total density and the phase of the condensate, the spin counterparts $\rho_s$ and $\phi_s$ are related to the density difference in the two components and to the relative phase of the two components.

In this basis, the action $S^{(2)}$ can be written as the sum of the actions in the density and spin channels
\begin{multline}
	S^{\rm (2)}=\int{dt\,d\mathbf{y}\sum_{r=d,s}\left\{
	\hbar\delta\hat{\rho}_r\frac{\partial\delta\hat{\phi}_r}{\partial t}+\right.}\\
	+{\left.\sum_{\substack{i=1}}^D\left[\frac{\hbar^2}{4m\rho_0}\left(\frac{1}{\ell_i(t)}\frac{\partial\delta\hat{\rho}_r}{\partial y_i}\right)^2+\frac{\hbar^2}{4m}\rho_0\left(\frac{1}{\ell_i(t)}\frac{\partial\delta\hat{\phi}_r}{\partial y_i}\right)^2\right]\right.}\\
	{\left.+\frac{g_{r}}{2\mathcal{V}(t)}\delta\hat{\rho}_r^2\right\}}.
\label{Eq:ActionScaled_2nd_Hom_DS}
\end{multline}
and the elementary excitations decouple into two branches of density and spin excitations. 
Here $g_d=g+g_{ab}$ and $g_s=g-g_{ab}$ indicate the strength of the effective atomic interactions involved in the density and the spin branches. Dynamical stability of the condensate imposes that both interaction constants are positive $g_{d,s}>0$. 

In the remaining of this section we derive the equations governing the dynamics of these fluctuations and discuss the features of their motion. For simplicity, we drop the subscript $d,s$ in the modes as the following considerations apply to both the types of excitations, and write in general $\delta\rho$ and $\delta\phi$. We keep the subscript only in $g_r$, so to distinguish the effective interaction strength seen by the two components.

The Euler-Lagrange equations for the fluctuations are obtained by minimizing the action in Eq.~\eqref{Eq:ActionScaled_2nd_Hom_DS} with respect to variations in $\delta\hat{\rho}$ and $\delta\hat{\phi}$:
\begin{align}
	&\hbar\frac{\partial\delta\hat{\phi}}{\partial t}+g_r\frac{\delta\hat{\rho}}{\mathcal{V}(t)}-\frac{\hbar^2}{2m\rho_0 }\sum_{\substack{i=1}}^D\left[\frac{1}{\ell_i^2(t)}\frac{\partial^2\delta\hat{\rho}}{\partial y_i^2}\right]=0,\label{Eq:FluctEuler}\\
	&\frac{\partial\delta\hat{\rho}}{\partial t}+\frac{\hbar}{2m}\rho_0\sum_{\substack{i=1}}^D\left[\frac{1}{\ell_i^2(t)}\frac{\partial^2\delta\hat{\phi}}{\partial y_i^2}\right]=0.
\label{Eq:FluctCont}
\end{align}
In the homogeneous limit here considered it is convenient to expand the fields in the plane wave basis and consider waves of given wavevector $\mathbf{q}$. Restricting to classical equation for this mode and inserting the normalization $\delta{\rho}_{\mathbf{q}}=(\rho_0/2)\varrho_{q}(t)\,e^{i\mathbf{q}\cdot\mathbf{y}}$ 
and $\delta{\phi}_{\mathbf{q}}=\varphi_q(t)\,e^{i\mathbf{q}\cdot\mathbf{y}}$ into Eqs.(\ref{Eq:FluctEuler}-\ref{Eq:FluctCont}),
we obtain the following equations for the time-dependent amplitudes $\varrho_q(t)$ and $\varphi_q(t)$:
\begin{align}
 &\hbar \dot{\varphi}_q+\left[\frac{g_r \rho_0}{2\mathcal{V}(t)}+\frac{\hbar^2\,\boldsymbol{\Pi}^2(t)}{4m}\right]\varrho_q=0,\label{Eq:FluctPhi_Q}\\
 &\dot{\varrho}_q-\frac{\hbar}{m}\boldsymbol{\Pi}^2(t)\varphi_q=0,\label{Eq:FluctRho_Q}
\end{align}
where we defined the time-dependent wave vector $\boldsymbol{\Pi}^2(t) \equiv \sum_{\substack{i=1}}^D q_i^2/\ell_i(t)^2$ rescaled by the condensate size. 

For a static configuration with a constant $\ell$, the solution of the equations of motion Eqs.~\eqref{Eq:FluctEuler} and \eqref{Eq:FluctCont} has the usual form $\exp(-i\Omega_r t)$ and recovers the well known Bogoliubov dispersion of a homogeneous, two-component condensate \cite{Abad2013}:
\begin{equation}
\Omega_{r}^2(q) = c_r^2\mathbf{q}^2 + \frac{\hbar^2 \mathbf{q}^4}{4m^2}.
\label{Eq:EqW}
\end{equation}
In this work, we focus on the most relevant $g>g_{ab}>0$ case where the effective interaction strength $g_s=g-g_{ab}$ experienced by the spin excitations is positive but lower than the one of the density excitations. In these conditions, stability is guaranteed but the frequencies of the spin modes are systematically smaller than one of the corresponding density modes. This feature will play a crucial role for our study of particle creation in the spin modes generated by a time-dependent density of the system.

\subsubsection{Role of dimensionality}
\label{sec:dimensionality}
Before proceeding, it is interesting to highlight some important features of the equations of motion Eqs.(\ref{Eq:FluctPhi_Q}-\ref{Eq:FluctRho_Q}). Consider for simplicity an isotropic system with equal trapping frequencies $\omega_i(t) \equiv \omega (t)$ and, thus, equal scaling factors $\ell_i(t) \equiv \ell (t)$. The time-dependent wave vector $\boldsymbol{\Pi}(t)$ then scales as $1/\ell(t)$, while the volume $\mathcal{V}(t)$ scales as $\ell^D(t)$, where we remind $D$ being the dimensionality of the system.

Depending on $D$, for an expanding condensate with $\ell(t)\to +\infty$, the square bracket in Eq.~\eqref{Eq:FluctPhi_Q} is eventually dominated by one or another term. For $D=1$, the second term accounting for the superluminal behaviour of the Bogoliubov dispersion decreases faster than the first term accounting for interactions, so that any given $\mathbf{q}$ mode eventually enters the sonic range. The situation is completely different in $D=3$, where the interaction term decreases faster and the mode eventually acquires a single-particle character. As it was pointed out in~\cite{Fedichev2004,Hauke-preheating}, the $D=2$ case is peculiar, as in this case the equations of motion recover the constant $\ell$ case upon a trivial rescaling of the time $d\tau=dt/\mathcal{V}$. As a result, in this dimensionality the time-dependence of $\ell(t)$ has no effect on the phase or density fluctuations.

Similar scaling arguments can be used to also highlight the peculiarities of our physically expanding condensate in comparison with the case where the expansion is simulated by means of a time-dependent collisional interaction strength $g_r(t)$ between atoms, e.g. by means of a Feshbach resonance \cite{Feshbach_Rev,Hung1213}. Even though this technique has been extensively exploited in the literature in order to study the effect of particle creation in an effective non-stationary spacetime for phonons in a BEC~\cite{Jain-Cosm-BEC-2007,Hauke-preheating}, some crucial points need highlighting: the volume $\mathcal{V}$ and the time-dependent wave vector $\boldsymbol{\Pi}$ do not change in time, so the kinetic energy of the mode under consideration remains constant. Since the interaction energy decreases instead in time via $g_r(t)$ during the analog expansion, any excitation mode will eventually acquire a single-particle character independently of the dimensionality.

Furthermore, the scaling factor is in this case fully pre-determined by the externally determined time-dependence of $g_r(t)$ and does not constitute an independent degree of freedom of the system. This poses serious problems if one aims at going beyond the physics of quantum fields on a pre-determined background and is interested to the coupled dynamics of the two.

\subsubsection{Mode freezing effect}
By deriving the motion equations Eqs.(\ref{Eq:FluctPhi_Q}-\ref{Eq:FluctRho_Q}) with respect to time and combining them, we can reformulate our dynamics in terms of second order differential equations that only involve the phase and the density fluctuations separately~\cite{Fischer2004,Eckel-Cosm-BEC-2018}:
\begin{align}
	&\ddot{\varphi}_q-\left[\frac{1}{\mathcal{C}_{r,q}^2(t)}\frac{d(\mathcal{C}_{r,q}^2(t))}{dt}\right]\dot{\varphi}_{q}+\omega_{r,q}^2(t)\varphi_q=0,\label{Eq:FluctPhi_Q2}\\
	&\ddot{\varrho}_q-\left[\frac{1}{\boldsymbol{\Pi}^2(t)}\frac{d(\boldsymbol{\Pi}^2(t))}{dt}\right]\dot{\varrho}_q+\omega_{r,q}^2(t)\varrho_q=0,\label{Eq:FluctRho_Q2}
\end{align}
where
\begin{align}
	\mathcal{C}_{r,q}^2(t) & \equiv c_r^2(t) + \frac{\hbar^2 \boldsymbol{\Pi}^2(t)}{4m^2},\\
	\omega_{r,q}^2 (t) & \equiv \mathcal{C}_{r,q}^2(t)\boldsymbol{\Pi}^2(t).
\label{Eq:QCW}
\end{align}
and we defined the time-dependent speed of sound 
\begin{equation}
c_r^2(t) \equiv {g_r\rho_0}/\left({2m\mathcal{V}(t)}\right)    
\end{equation}
which is induced by the dynamical evolution of the density component of the system via the volume scaling factor $\mathcal{V}(t)$. 
The Eqs.~\eqref{Eq:FluctPhi_Q2} and \eqref{Eq:FluctRho_Q2} show that, in the case of an expanding condensate, each mode, as seen in co-moving coordinates, undergoes the dynamics of a damped harmonic oscillator with a time-dependent frequency. The effective damping experienced by the modes appears because of the variation in size of the system, and is the analogous of the cosmological \emph{Hubble friction} that originates in an expanding Universe \cite{parker_toms_2009}. 

Even though the physical picture of the Hubble friction is a useful tool to intuitively understand the physics, some peculiar features are worth being pointed out. First, the effective friction experienced by the phase and density fluctuations in respectively \eqref{Eq:FluctPhi_Q2} and \eqref{Eq:FluctRho_Q2} seem to have different physical origins. For the density fluctuations, the effective friction is related to the redshift of the modes consequent to the variation in size of the system itself. For the phase, it appears via a time dependence in the speed of sound of the modes, which in turn depends (in the hydrodynamic limit) on the density of the system as well as on the interaction constant $g$. 

The evolution given by 
Eqs.(\ref{Eq:FluctPhi_Q2}-\ref{Eq:FluctRho_Q2}) is illustrated in Fig.~\ref{Fig:2} for the case of an expanding system. 
For simplicity we focus again on an isotropic system with equal trapping frequencies $\omega_i(t) \equiv \omega (t)$ and $\ell_i(t) \equiv \ell (t)$. Also, let us define the Hubble parameter $H(t) \equiv \dot{\ell}(t)/\ell(t)$. This is proportional the friction appearing in Eqs.~\eqref{Eq:FluctPhi_Q2} and \eqref{Eq:FluctRho_Q2}. It is constant in the case of an exponential expansion: $\ell(t)\sim e^{H t}$, while $H(t)\sim 1/t$ for the linearly expanding system: $\ell(t)\sim t$.
As a specific example of the general physics, in the Figure we show the time evolution of the density and phase components of a spin mode of frequency $\omega_s/\omega_0 = 50$, obtained by numerically solving the equations of motion for the case of an exponentially expanding one-dimensional condensate.

As shown in the Figure, the dynamical evolution of the Bogoliubov modes goes through three different stages, that arise as a result of the competition between the different time scales provided by the mode frequency and the effective friction in the motion equations. At the early times of the expansion [indicated as (I) in the Figure], when the expansion rate is still negligible respect to the mode natural frequency, the mode evolves as an almost free harmonic oscillator. As the expansion proceeds [temporal region (II) in the Figure], the value of the frequency decreases because of the combined effect of the redshift of the wavelengths and of the reduced density that result in a lower value of the sound speed. This change in frequency is also responsible for a redistribution of the amplitude between the density and phase components that is visible in the Figure. This effect can be derived from the equation relating the time evolution of the amplitudes of the phase and density components of the modes, which is readily obtained from Eqs.~(\ref{Eq:FluctPhi_Q}-\ref{Eq:FluctRho_Q}) as
\begin{multline}
  \frac{\hbar^2}{m} \frac{d|\varphi_q|^2}{dt} +\\ \frac{1}{\mathcal{V}(t)\boldsymbol{\Pi}^2(t)}\left[\frac{g_r}{2}+\frac{\hbar^2}{4m}\frac{\mathcal{V}(t)}{\varrho_0}\boldsymbol{\Pi}^2(t)\right]\frac{d|\varrho_q|^2}{dt}=0.
    \label{Eq:AmplEvol}
\end{multline}
Since the time dependent coefficients in Eq.~\eqref{Eq:AmplEvol} are positive, we deduce that the variation of the amplitude of the phase and density components is opposite in sign. 
At the time when the oscillation frequency becomes comparable to the expansion rate $\omega_{r,q}(t)\approx H(t)$, the effective damping starts to strongly affect the dynamics that turns into an over-damped regime [temporal region (III) in the Figure] analogous to the dynamics of a mass attached to a spring that oscillates immersed in a viscous medium. Both the values of the elastic constant of the spring and of the viscosity goes to zero over time: since the former decays faster than the latter, the mode amplitude tends to a finite constant value in the long time limit.

A similar phenomenology occurs in Cosmology when the wavelength of a mode crosses the so-called  Hubble radius $R_H(t)$, that is defined in terms of the sound speed $c_r(t)$ and the Hubble parameter $H(t)$ as $R_H (t) \equiv c_r(t)/H(t)$. Physically this represents the distance between two points moving away from each other with luminal velocity. This interpretation is readily demonstrated by posing the relative physical velocity $v$ between two points
\[v = \frac{d}{dt} |\mathbf{x}| = \frac{d}{dt}\left(\ell(t)|\mathbf{y}|\right) = \frac{\dot{\ell}(t)}{\ell(t)}|\mathbf{x}|=H(t)|\mathbf{x}|\]
equal to $c_r(t)$. The Hubble radius is a local quantity (it is defined at each time instant) and has not to be confused with the (past and future) cosmological horizons, that are global features of spacetime instead \cite{parker_toms_2009}. 
In this late stage of the evolution,  the mode frequencies go to zero faster than $H(t)$ and the amplitudes display an over-damped behaviour towards a finite-valued long-time limit. In the Cosmological literature, this phenomenology goes under the name of \emph{mode freezing}. 

This mode freezing is generic to all dimensions and can be reconciled with our previous analysis of the $D=2$ case where the scaling arguments predict the absence of evolution of the density and the phase. To this purpose, one need to note that the rescaled time $\tau$ has a finite limit for a physical time $t\to \infty$ for any $D$. Since the modes keep oscillating at the Bogoliubov frequency in the rescaled temporal variable $\tau$, it is immediate to understand why the mode amplitudes shown in the Figure tend to a constant value for $t\to +\infty$.

As a final remark, it is useful to comment on the physical nature of the Hubble friction. Since our evolution is a purely conservative one, the Hubble friction is only apparent and is not associated to any real dissipation process. In particular, if one considers the combination of suitably tuned expansion and  contraction stages, the system can be brought back to its initial quantum state without inserting any additional noise. On one hand, this can be be understood as the sign of the friction being reversed when expansion is replaced by contraction, leading to an effective amplification. On the other hand, the evolution of our system differs from the one of a generic quantum system experiencing a sequence of dissipation and reamplification stages, as in this case the overall process would unavoidably introduce some extra noise. These remarks highlight the necessity of using the expression {\em mode freezing} with due care.

\subsection{Effective Hamiltonian}
In this subsection we derive the quantum mechanical energy operator for the Bogoliubov excitations in the general case of a non-stationary condensate. As a first step towards this objective, we derive first the scalar product of the corresponding field theory from first principles.

\subsubsection{Scalar product}
\noindent Given the action in Eq.~\eqref{Eq:ActionScaled_2nd_Hom_DS}, the scalar product is defined as the space integral of the time-component of the conserved $(D+1)$-current $J^\alpha$ resulting from the global phase invariance of the Lagrangian. The explicit expression for such a scalar product is obtained by first generalizing the Lagrangian of the theory to the case of complex $\varphi_r$ and $\varrho_r$ fields since, in the homogeneous limit we are considering, we are expanding the phase and density fields in plane waves. The conserved current is a classical concept, so we work with classical fields in this section. The first term in Eq.~\eqref{Eq:ActionScaled_2nd_Hom_DS} can be generalised as
\begin{equation}
\frac{\hbar}{2}\left(\delta\rho_{r}^*\frac{\delta\partial\phi_{r}}{\partial t}+\delta\rho_{r}\frac{\partial\delta\phi_{r}^*}{\partial t}\right),
\label{Eq:CpxLagrangian}
\end{equation}
having opportunely symmetrized the time derivative between the density and phase fields. A similar procedure can be applied to the other terms of the Lagrangian. The resulting complex Lagrangian is invariant under the transformations
\begin{subequations}
\begin{align}
\delta\rho_{r}\rightarrow\delta\rho_{r} e^{i\epsilon}\approx\delta\rho_{r}(1+i\epsilon),\\
\delta\phi_{r}\rightarrow\delta\phi_{r} e^{i\epsilon}\approx\delta\phi_{r}(1+i\epsilon),
\end{align}
\end{subequations}
where $\epsilon$ is an arbitrary infinitesimal phase. The conservation law is deduced from the Noether theorem \cite{Peskin-book}, and is written as
\begin{equation}
	\partial_\alpha J_r^\alpha=0,
\label{Eq:Current}
\end{equation}
where
\begin{equation}
	J_r^\alpha=\sum_{\substack{\sigma=\delta\phi,\delta\rho}}\left[\frac{\partial\mathcal{L}}{\partial\left(\partial_\alpha\sigma_r\right)}\delta\left(\partial_\alpha\sigma_r\right)+\frac{\partial\mathcal{L}}{\partial\left(\partial_\alpha\sigma_r^*\right)}\delta\left(\partial_\alpha\sigma_r^*\right)\right],
\end{equation}
is the conserved $(D+1)-$current for each of the two components. The Eq.~\eqref{Eq:Current} is a continuity equation. By integrating it over the spatial volume, and considering field variations that vanish at the spatial boundaries, we obtain
\begin{equation}
\partial_t\left(\int{d\mathbf{y} J_r^0}\right)=0,
\end{equation}
provided $\delta\phi_r$ and $\delta\rho_r$ are solution of the field equations in Eqs.~\eqref{Eq:FluctEuler} and \eqref{Eq:FluctCont}. The spatial integral of the time-component $J_r^0$ of the current is thus constant. In the case of a complex field theory whose quanta are distinguishable particles with opposite charge, $J_r^0$ has the physical meaning of total charge in the system. In the case of a condensate instead the fields $\varphi_r$ and $\varrho_r$ are real-valued and we have a single type of particle that is the phonon, and we can assign to the conserved quantity the meaning of a scalar product. We thus have
\begin{align}
&\int{d\mathbf{y}\,J_r^0}=\\
&=\frac{i\hbar}{2}\epsilon\int{d\mathbf{y}\left(\delta\rho_r^*\delta\phi_r-\delta\rho_r\delta\phi_r^*\right)}=\text{const.}\quad.
\end{align}
Since the Lagrangian is quadratic, this conservation law holds for each couple of density and phase modes, individually. For a reason that will be clear in the next sections, we chose the arbitrary phase $\epsilon$ in such a way that the scalar product takes the explicit form
\begin{equation}
\left(\varphi_{r,q},\varrho_{r,q}\right)\equiv \varphi_{r,q}^*\varrho_{r,q}-\varrho_{r,q}^*\varphi_{r,q}=i/N,
\label{Eq:ScalarProd}
\end{equation}
where we used the notation introduced in the previous section for the amplitude of the modes in the homogeneous system, and we indicated by $N=\int{d\mathbf{y}\,\rho_0(\mathbf{y})}$ the total number of particles in the system.

\subsubsection{Time-dependent Hamiltonian}
\noindent The energy of the excitations is obtained by integrating over the spatial domain the term of the energy density in Eq.~\eqref{Eq:EnFunc} that is of second order in the quantum fluctuations. By working in the spin and density basis, and by using the scaled quantities and co-moving coordinates, this energy can be written as
\begin{multline}
\hat{E}^{(2)}=\int{d\mathbf{y}\sum_{r=d,s} \left\{\sum_{i=1}^D\left[ \frac{\hbar^2}{2m\rho_0} \left(\frac{1}{\ell_i^2(t)}\frac{\partial^2\delta\hat{\rho}_r}{\partial y_i^2}\right)^2   \right.\right.}\\
\left.\left. +\frac{\hbar^2}{2m} \rho_0 \left(\frac{1}{\ell_i^2(t)}\frac{\partial^2\delta\hat{\phi}_r}{\partial y_i^2}\right)^2 \right] + \frac{g_r}{\mathcal{V}(t)} \delta\hat{\rho}_r^2 \right\}.
\end{multline}
By using the Eqs.~\eqref{Eq:FluctEuler} and \eqref{Eq:FluctCont}, this reduces to the simple form
\begin{equation}
\hat{E}^{(2)}=\int{d\mathbf{y}\sum_{r=d,s} \hbar\left(\delta\hat{\phi}_r \frac{d\delta\hat{\rho}_r}{dt} - \delta\hat{\rho}_r \frac{d\delta\hat{\phi}_r}{dt}\right).}
\end{equation}
By using the expansion of the quantum fluctuations in terms of the Bogoliubov modes: 
\begin{align}
	\delta\hat{\rho}_r &= \frac{\rho_0}{2}\sum_{\mathbf{q}}\left(\varrho_{r,q} e^{i\mathbf{q}\cdot\mathbf{y}} \hat{b}_\mathbf{q} + \varrho_{r,q}^* e^{-i\mathbf{q}\cdot\mathbf{y}}\hat{b}_\mathbf{q}^\dag\right), \\
	\delta\hat{\phi}_r &= \sum_{\mathbf{q}}\left(\varphi_{r,q} e^{i\mathbf{q}\cdot\mathbf{y}} \hat{b}_\mathbf{q} + \varphi_{r,q}^* e^{-i\mathbf{q}\cdot\mathbf{y}}\hat{b}_\mathbf{q}^\dag\right), 
\end{align}
this energy operator can be expanded as
\begin{widetext}
\begin{equation}
\frac{\hat{E}^{(2)}}{N}=\frac{\hbar}{2}\sum_{r,\mathbf{q}}\left(W\left[\varphi_{r,q},\varrho_{r,q}^*\right]+\left(W\left[\varphi_{r,q},\varrho_{r,q}^*\right]\right)^*\right)\hat{b}_{r,\mathbf{q}}^\dag\hat{b}_{r,\mathbf{q}}+\frac{\hbar}{2}\sum_{r,\mathbf{q}}{W\left[\varphi_{r,q},\varrho_{r,q}^*\right]}+\frac{\hbar}{2}\sum_{r,\mathbf{q}}\left(W\left[\varphi_{r,q}^*,\varrho_{r,q}^*\right]\hat{b}_{r,\mathbf{q}}^\dag\hat{b}_{r,-\mathbf{q}}^\dag+h.c.\right).
\label{Eq:E2}
\end{equation}
\end{widetext}
Here we defined the Wronskian $W\left[\varphi_{r,q},\varrho_{r,q}\right]\equiv \varphi_{r,q}\dot{\varrho}_{r,q}-\dot{\varphi}_{r,q}\varrho_{r,q}$ and used the relation $W\left[\varrho_{r,q},\varphi_{r,q}^*\right]=-(W\left[\varphi_{r,q},\varrho_{r,q}^*\right])^*$. The first term in Eq.~\eqref{Eq:E2} represents the energy carried by the quasi-particles that populate each of the Bogoliubov modes. The second term accounts instead for the zero-point contribution of the vacuum. If different from zero, the last term in Eq.~\eqref{Eq:E2} accounts for a process of squeezing, and thus the creation of entangled pairs of quasi-particles with opposite momenta. 

At equilibrium this term of course has to be zero. In such conditions the frequencies of the modes are well defined, and they take the form (see Eqs.~\eqref{Eq:EqW}, \eqref{Eq:FluctPhi_Q2} and \eqref{Eq:FluctRho_Q2}): $\varphi_{r,q}=\bar{\varphi}_{r,q}e^{-i\Omega_{r}(q) t}$, $\varrho_{r,q}=\bar{\varrho}_{r,q}e^{-i\Omega_{r}(q) t}$ (with $\bar{\varphi}_{r,q}$, $\bar{\varrho}_{r,q}$ complex constants). By substituting these expressions into the definition of the Wronskian, we obtain
\begin{align}
W\left[\varphi_{r,q},\varrho_{r,q}^*\right]&=2i\Omega_{r}(q)\bar{\varphi}_{r,q}\bar{\varrho}_{r,q}^*,\label{Eq:W1}\\
W\left[\varphi_{r,q},\varrho_{r,q}\right]&=0\label{Eq:W2},
\end{align}
so that the energy can be rewritten as
\begin{multline}
\frac{E^{(2)}}{N}=\hbar\sum_{r,\mathbf{q}}\Omega_{r}(q)\left[i \left(\bar{\varphi}_{r,q}\bar{\varrho}_{r,q}^*-\bar{\varphi}_{r,q}^*\bar{\varrho}_{r,q}\right) \right]\hat{b}_{r,\mathbf{q}}^\dag\hat{b}_{r,\mathbf{q}}\\
+\frac{\hbar}{2}\sum_{r,\mathbf{q}}{2i\Omega_{r}(q)\bar{\varphi}_{r,q}\bar{\varrho}_{r,q}^*}.
\label{Eq:E2_Static}
\end{multline}
From the Eq.~\eqref{Eq:E2_Static} we infer that the theory has a particle interpretation if
\[N\left(\bar{\varphi}_{r,q}^*\bar{\varrho}_{r,q}-\bar{\varphi}_{r,q}\bar{\varrho}_{r,q}^*\right)=i,\]
that is the scalar product defined in Eq.~\eqref{Eq:ScalarProd}. Also note that, in the static configuration, $N\bar{\varphi}_q^*\bar{\varrho}_q=i/2$. The energy is conserved in this case as expected, and takes the standard form at equilibrium:
\begin{equation}
E^{(2)}=\sum_{r,\mathbf{q}}\hbar\Omega_r (q) \left(\hat{b}_{r,\mathbf{q}}^\dag\hat{b}_{r,\mathbf{q}}+\frac{1}{2}\right).
\label{Eq:E2_Static_Std}
\end{equation}
In the general time-dependent case, the Eqs.~\eqref{Eq:W1},\eqref{Eq:W2} are not verified, and $W\left[\varphi_{r,q},\varrho_{r,q}\right]$ is different from zero. This means that the energy is not conserved and pairs of entangled particles with opposite momenta are created out of the initial vacuum state. 

We should mention here that the non conservation of the energy is a consequence of the time-dependence of the background underlying our quantum field. This highlights the fact that the Bogoliubov theory adopted here only provides a partial description of the system in terms of a time-dependent Hamiltonian. In particular, this model is not self-consistent as it does not take into account the effects of the back-reaction of the quantum fluctuations onto the mean-field component. A more sophisticated theory solving this difficulty will appear in a forthcoming work~\cite{new}.

\section{Thomas-Fermi non-stationary one-dimensional condensate\label{Sec:Homog}}

\noindent 
In the previous section we developed the theory that describes the dynamics of the Bogoliubov excitations in a non-stationary, two-component condensate. We developed this model by working in the TF limit, in which the interaction energy is the predominant energy scale in the mean-field description of the system, and we considered the spatial region close to the centre of the trapping potential in order to justify our assumption of homogeneous density. In more formal terms, the density can be approximated as homogeneous when it changes over a length scale that is much longer compared to the characteristic microscopic length scale of the condensate. The former is provided by the TF radius $R_{\rm TF}(t)\equiv [2\mu/m\omega^2(t)]^{1/2}$, which gives the spatial extension of the condensate, while the latter is provided by the healing length $\xi_r = \{\hbar^2/[m g_r \rho_0(\mathbf{x}=0)]\}^{1/2}$. The definition of the healing length is not unique for the two-component system, as collective modes in the spin and density branches experience a different effective interaction strength. Since one typically has $g_s<g_d$, the validity condition $\xi_r/R_{\rm TF}(t)\ll 1$ for the constant density approximation is more easily verified for the density modes rather than the spin modes. 

%In our investigation of particle production out of the spin vacuum, we are actually interested in working with relatively small values for the sound speed $c_s$ and thus for the interaction strength $g_s$. Lower the value of $c_s$ the stronger is in fact the modulation of the background density experienced by the spin modes, resulting in a stronger signal from the amplification of the zero-point fluctuations. This requirement makes however more difficult to verify the $\xi_s/R_{\rm TF}(t)\ll 1$ condition, unless we work with very large systems. However, in any case, the homogeneous theory we developed is able to describe with high fidelity the dynamics of the density modes and provides a reliable qualitatively description to the dynamics of the spin excitations. To overcome the limitations of the model, a full numerical analysis of the inhomogeneous system is discussed in the next section.

%\noindent In this section we use the homogeneous model to study the particle creation generated by a condensate that is expanding and oscillating. For simplicity, and without affecting the generality of the following arguments, we consider the one-dimensional configuration.

\begin{figure*}[!t]
\centering
{\subfigure
{\includegraphics[width=0.4\textwidth]{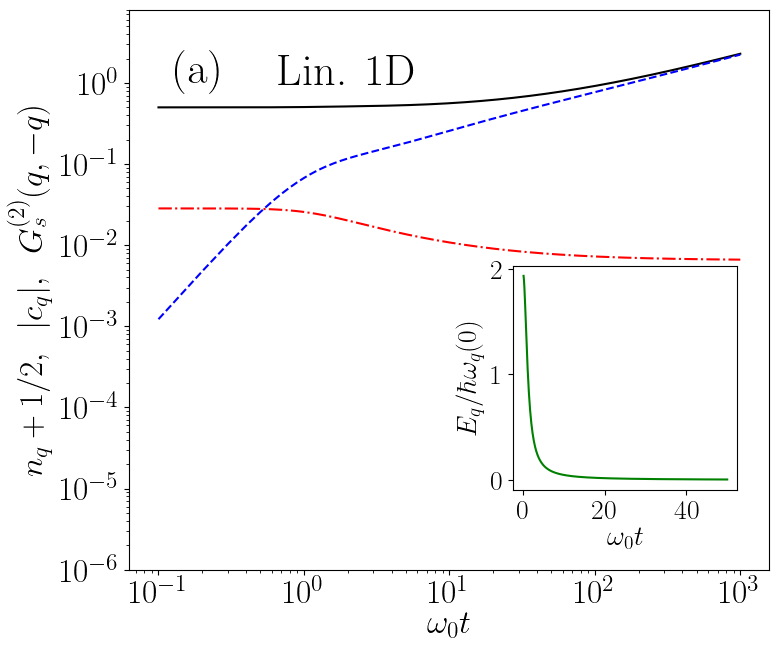}}}
{\subfigure
{\includegraphics[width=0.4\textwidth]{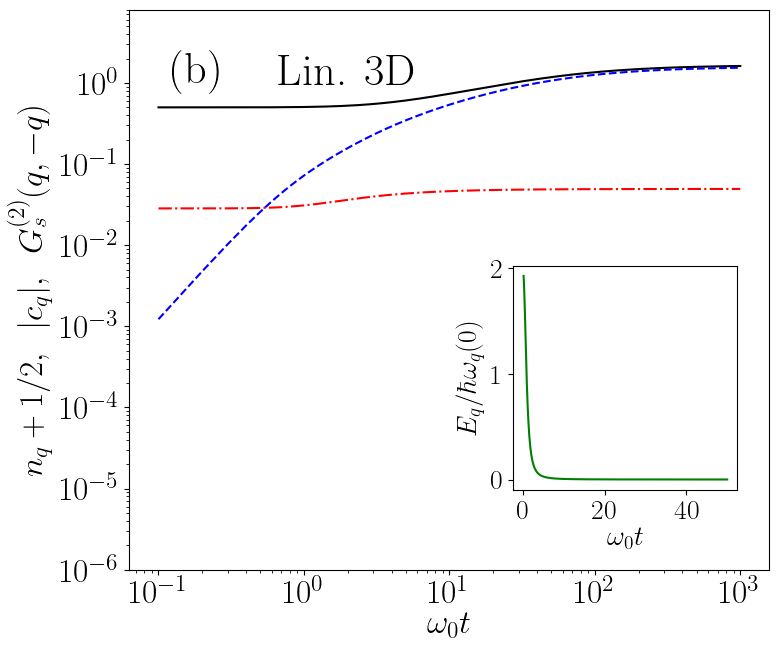}}}\\
{\subfigure
{\includegraphics[width=0.4\textwidth]{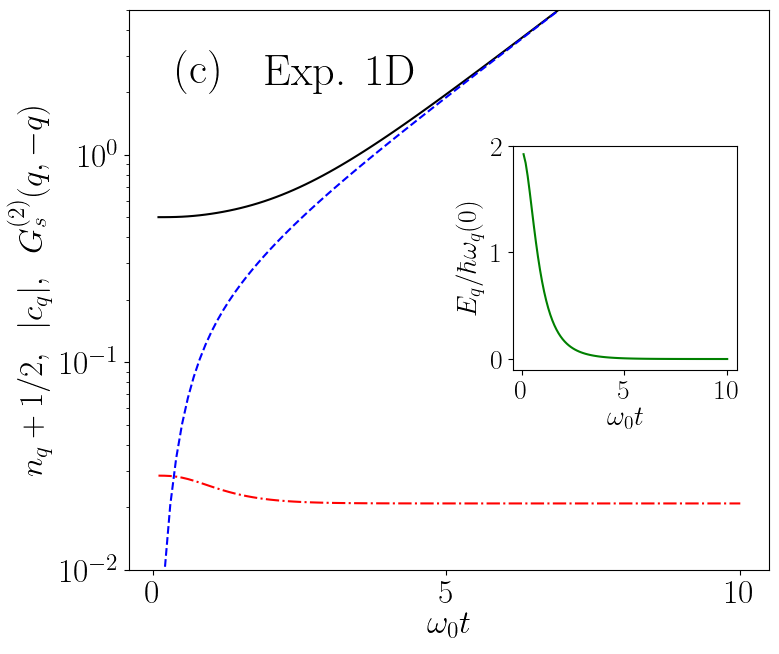}}}
{\subfigure
{\includegraphics[width=0.4\textwidth]{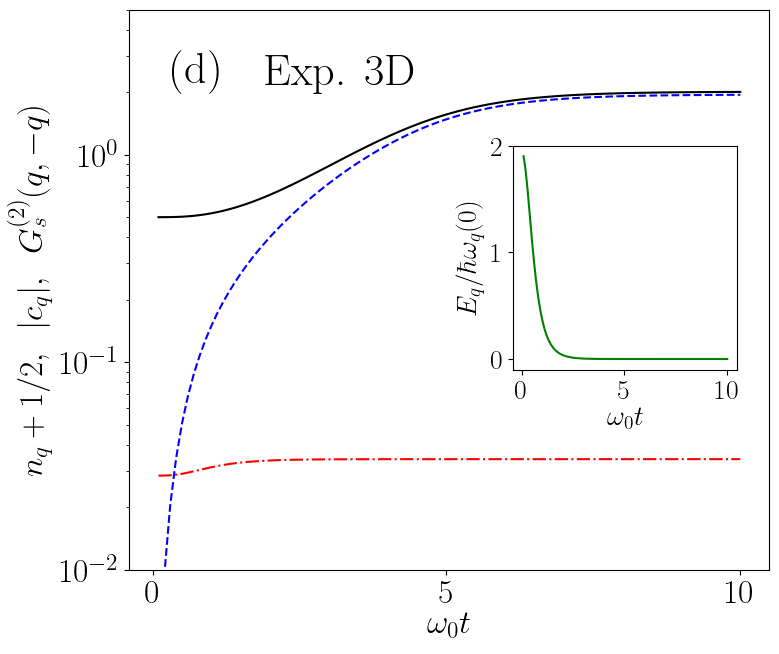}}}\\
{\subfigure
{\includegraphics[width=0.4\textwidth]{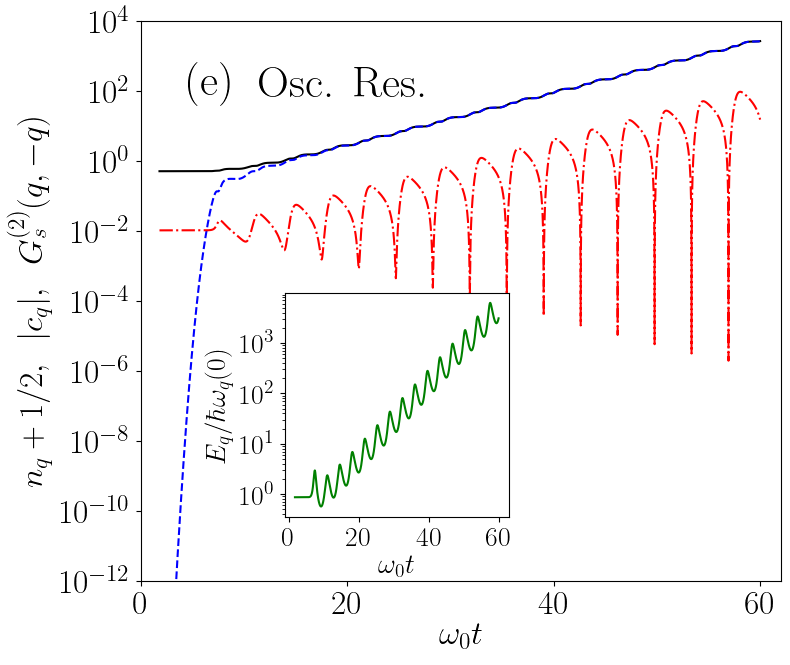}}}
{\subfigure
{\includegraphics[width=0.4\textwidth]{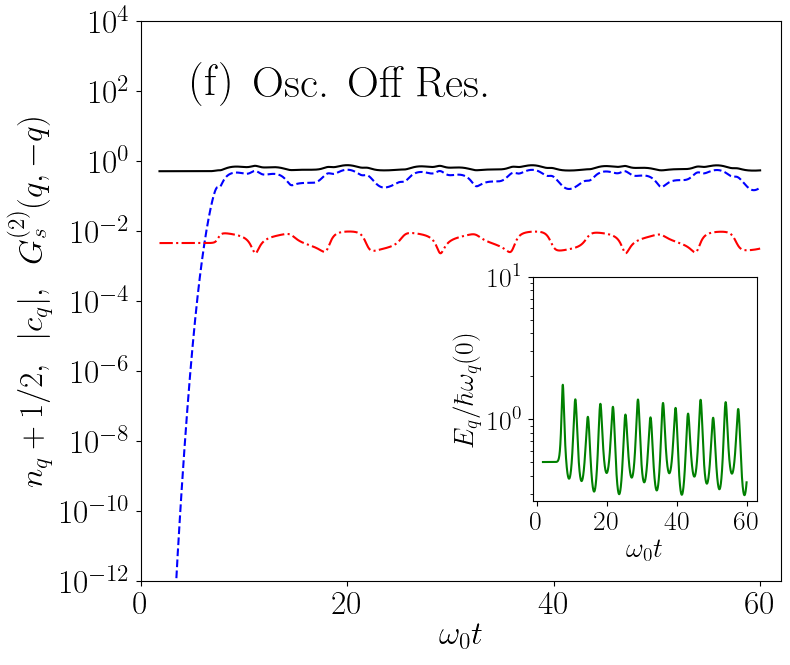}}}
\caption{Panels (a-d) (upper and middle row): time evolution of the number of excitations $n_q$, including the initial vacuum fluctuations,  $n_q+1/2\equiv W[\varphi_q,\varrho_q^*]/(2\omega_q(t))$ (solid black lines), of the modulus of the anomalous correlations $c_q \equiv W[\varphi_q,\varrho_q]/(2\omega_q(t))$ (dashed blue lines) and of the correlation functions of the density fluctuations $G_2^{(2)}(q,-q)$ (dot-dashed red line), generated in the spin mode of frequency $\omega_{s,1}/\omega_0=0.5$, of a condensate that is linearly (a,c) and exponentially (b,d) expanding in one (a,c) and three (b,d) dimensions. The inset shows the time evolution of the energy of the mode, which goes to zero because of the freezing effect. Panels (e,f) (bottom row): same quantities for a condensate that oscillates in the breathing density mode. The two panels refer respectively to a resonant case with $\omega_{s,2} \approx \omega_{d,2}/2\approx\sqrt{3}/2$ (e) and to a non-resonant case with $\omega_{s,1}/\omega_0 = 0.5$. In all panels, we have taken $g_d/g_s = 4$. %The equal value $n_q = c_q$ attained at the later time in panels (a-c) is a clear effect of the squeezing of the modes.
}
\label{Fig:1}
\end{figure*}

\subsection{Particle creation and correlations}
\noindent The coupled dynamics of the background condensate and the quantum fluctuations is governed by the set of Eqs.~\eqref{Eq:B_Evol}, \eqref{Eq:FluctPhi_Q2} and \eqref{Eq:FluctRho_Q2}. The homogeneous spectrum, in Eq.~\eqref{Eq:QCW} reproduces in the long wavelength limit the TF spectrum $[\omega_{r,n}^{\rm TF}/\omega_0]^2 = \frac{g_r}{g_d}\left[\frac{n}{2}\left(n +1\right)\right]$, provided we take wave vectors of the form $q_n(t)=n/R_{\rm TF}(t)$ $(n = 1,2,...)$ and add the constant term $c_r^2(t)q_n(t)/R_{\rm TF}(t)$. With this ad-hoc amendments it reads as
\begin{align}
	\omega_{r,n}^2(t) &= \frac{c_r^2(t)q_n(t)}{R_{\rm TF}(t)} + c_r^2(t)q_n^2(t)+\mathcal{O}[q_n^4(t)],\nonumber\\
	&=\frac{\left[\omega_{r,n}^{\rm TF}\right]^2}{\mathcal{V}(t)\ell^2(t)}+\mathcal{O}[q_n^4(t)].
\label{Eq:Wmod}
\end{align}

By using this expressions for the frequency in Eqs.~\eqref{Eq:FluctPhi_Q2} and \eqref{Eq:FluctRho_Q2}, together with the values for the wave-vectors given above, we are thus able to simulate the dynamics of the Bogoliubov excitations on top of a TF condensate.

%\newpage
The Eqs,~\eqref{Eq:FluctPhi_Q2} and \eqref{Eq:FluctRho_Q2} are thus solved, given the initial conditions provided by the mode functions at equilibrium, that read
\begin{subequations}
\begin{align}
	\varrho_{r,q}(t)&=\sqrt{\frac{\epsilon_{q,0}}{\hbar\omega_{r,q}}}\,e^{-i\Omega_{r,q} t},\\
	 \varphi_{r,q}(t)&=-\frac{i}{2}\sqrt{\frac{\hbar\omega_{r,q}}{\epsilon_{q,0}}}e^{-i\Omega_{r,q} t},
\end{align}
\end{subequations}
with $\epsilon_{q,0}^2={\hbar^2 q^2}/{2m}$. 

We consider the two configurations of an expanding condensate or of an oscillating condensate in its breathing density mode. In the former case, a linear or exponential expansion is implemented by switching-off or reverting the sign of the trapping potential, respectively. The oscillating condensate is instead implemented by perturbing the trapping potential in order to excite the density breathing mode of the system. To this aim, we consider a sudden modulation of the trapping frequency with a temporally-localized form, $\omega(t)/\omega_0=1+A\exp(-(t-t_0)^2/(2\sigma^2))$. Here $A$ is the amplitude of the modulation, $t_0$ is time instant at which it takes place, while $\sigma$ determines its duration. This perturbation sets the condensate in motion mostly in its breathing mode~\cite{Stringari_BEC}, with the expansion parameter $\ell(t)$ periodically oscillating around its equilibrium value $\ell(0)=1$. 

Because of the periodicity of the oscillations, a resonant parametric coupling between the density and spin branches is then triggered, which involves modes whose frequencies are related as $\omega_{d,n} = 2 \omega_{s,m}$. In the case of the breathing oscillations here considered $(n=2)$, and limiting to the long wavelength regime, the spin modes verifying the resonance condition are the ones for which
\[m(m+1) = \frac{3}{2}\frac{g_d}{g_s}.\]
This means that, depending on the value of the ratio $g_d/g_s$, a different spin mode is resonant with the breathing density mode.

In the case of an expanding system, we have seen in Sec.~\ref{sec:dimensionality} that the Bogoliubov modes ultimately freeze, attaining a constant value. This is due to the fact that the frequency of the modes goes to zero faster than the Hubble parameter, yet with different laws depending on the dimensionality. 

Since the energy of the quantum fluctuations is proportional to the Wronskian $W[\varphi_{r,q},\varrho^*_{r,q}]$ calculated for the phase and the density amplitudes of the modes, the mode freezing effect implies that the energy of each mode eventually goes to zero at late times of the expansion. This physically means that the expansion drives the fluctuations towards a final cold state, as in the case of a monotonically expanding Universe.
At a closer look, however, one notes that the time-dependent background parametrically amplifies the zero-point fluctuations in the spin modes, and pairs of entangled (quasi-)particles are created out of the Bogoliubov vacuum. The number of particles $n_q$ created in a certain mode out of the vacuum is obtained by evaluating the quantity $n_q+1/2\equiv W[\varphi_q,\varrho_q^*]/(2\omega_q(t))$, which comprises also the initial vacuum fluctuations in the mode. These particles are created in a \emph{squeezed} state, as entangled pairs with opposite momenta. The build up of quantum correlations is witnessed by the quantity $c_q = W[\varphi_q,\varrho_q]/(2\omega_q(t))$. The results reported in Figs.~\ref{Fig:1}(a-d) clearly show that the late state of a spin mode in an expanding condensate is squeezed, as $n_q = c_q$. The fact that $n_q,c_q$ saturate to a finite value in $D=3$ whereas they keep growing in $D=1$ can be understood in terms of the single-particle (sonic) nature of the $q$ mode at late times in $D=3$ ($D=1$). 

A similar particle creation effect takes place in the oscillating case, if a resonant mode is considered. Figs.~\ref{Fig:1}(e,f) show the time evolution of the number of excitations $n_q$ and the correlations $c_q$ in modes that are either resonant or off-resonant with the (halved) frequency of density oscillations: In the former case, the number of particles that populate the mode grows exponentially, while it remains almost unaltered in the latter case.

Rather than looking at the number of excitations $n_q$ or at the correlations $c_q$, it is often more convenient in actual experiments to consider the correlation function of density fluctuations: $G_s^{(2)}(q,-q) \equiv \left<\delta\hat{\rho}_s(q) \delta\hat{\rho}_s(-q)\right>$. Given the system initially in the vacuum state, this reduces to $G_s^{(2)}(q,-q) = \rho_0^2|\varrho_{s,q}|^2/4$. In all panels of the same Figure, we plot as red dot-dashed lines the time-evolution of the component of spin density fluctuations at wavevector $q$ that results from the excitation of the Bogoliubov mode at this wavevector. 

On one hand, no marked feature is visible for an expanding condensate. In this case, the creation of quasi-particles gets in fact intertwined with the change in the collective vs. single-particle character of the $q$ mode. This is visible as a difference between the $D=1$ and $D=3$ cases: In agreement with our discussion in Sec.~\ref{sec:dimensionality}, in $D=1$ [panels (a,c)] the mode eventually becomes a collective excitation with a mostly phase character, so the spin-density fluctuations get suppressed. In $D=3$ [panels (b-d)], instead, the mode eventually gets a  single-particle mode character recovering a sizable amplitude of spin-density fluctuations; the fact that the long-time limit does not reach the value $1/4$ of the vacuum state of single-particle modes is a signature of the squeezing associated to the particle creation process.
In all dimensions $D$, the constant and non-oscillating late-time value of the spin density fluctuations is a signature of the mode freezing effect. 

On the other hand, a clearly visible signal is found in every dimension for a resonantly oscillating condensate [panel (e)], which looks very promising in view of experiments. The large contrast of the oscillations in the spin density fluctuations is a signature of squeezing effects, which have been predicted to lead to non-separable behaviours~\cite{Parentani-NonSeparability,Robertson-PRD-PreHeatAn-2019}.

\begin{figure}[htbp]
\centering
\includegraphics[width=0.45\textwidth]{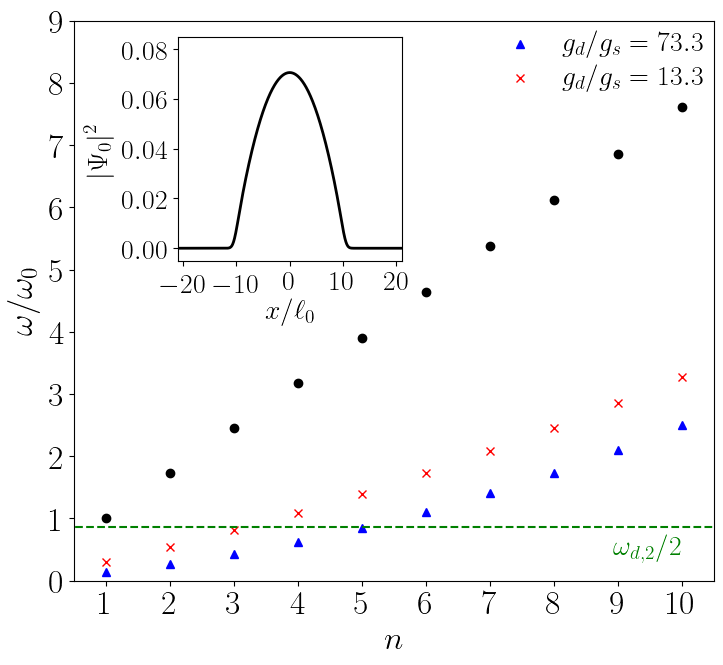}
\caption{Bogoliubov spectrum of the density (black dots) and spin (red and blue markers) modes in a one-dimensional, harmonically trapped two-component condensate of chemical potential $\mu/\omega_0=28.25$. We notice that the spin modes of quantum number $n=3$ (for $g_d/g_s=13.3$, red marker) and $n=5$ (for $g_d/g_s=73.3$, blue marker) are close to resonance with the breathing mode (n=2) in the density branch of the excitations. We reported in the inset the density profile of the condensate.}
\label{Fig:3}
\end{figure}

\section{Inhomogeneous non-stationary one-dimensional condensate\label{Sec:InHom}}

In order to further validate the predictions of the theoretical model presented in the previous section and based on a homogeneous system approximation, we report now a full numerical study for the particle creation in the inhomogeneous system. We pursue this analysis by using the same physical configurations previously discussed, that are the expanding and oscillating systems. 

In the perspective of the experimental investigation of this physics, we focus here on the (connected component
of the) density  and spin $G_{d,s}^{(2)}(x,x')$ correlation functions. Correlation functions have revealed to be particularly useful in order to detect the weak signal arising from the amplification of the zero-point fluctuations of a quantum field in condensed matter analog models \cite{steinhauer2016,steinhauer2019,steinhauer2021,steinhauer2021-Light}. In our two-component case, these are defined as
\begin{align}
	G_{d}^{(2)}(x,x') &= \left<\hat{\Rho}(x)\hat{\Rho}(x')\right> - \left<\hat{\Rho}(x)\right> \left<\hat{\Rho}(x')\right>,\label{Eq:G2_n}\\
	G_{s}^{(2)}(x,x') &= \left<\hat{S}(x)\hat{S}(x')\right> - \left<\hat{S}(x)\right> \left<\hat{S}(x')\right>,
\label{Eq:G2_s}
\end{align}
where $\hat{\Rho} = \hat{\Rho}_a + \hat{\Rho}_b$ is the total density operator, while $\hat{S} = \hat{\Rho}_a - \hat{\Rho}_b$ is the spin density operator that accounts for the excess of particles in one species compared to the other. For numerical ease we work now with the complex field operators $\hat{\Psi}_j$ $(j=a,b)$ rather than with the (real) density and phase fields. Within this formalism, the fields can be split according to the Bogoliubov prescription as $\hat{\Psi}_j = \Psi_{j,0} + \delta\hat{\Psi}_j$. The first term accounts for the mean-field component, that in our symmetric configuration is equal for both the atomic components and reads: $ \Psi_{0,a} = \Psi_{0,b} =\Psi_0 /\sqrt{2}$. The quantum component can be conveniently written in the spin and density basis. In terms of the standard $u,v$ eigenfunctions, this reads \cite{Stringari_BEC}:
\begin{equation}
    \delta\hat{\Psi}_r = \sum_{n\in (+)}\left(u_{r,n} \hat{b}_{r,n}+v_{r,n}^*\hat{b}_{r,n}^\dag\right),
\label{Eq:Psi_uv_NotHom}
\end{equation}
in which the sum runs over the positive norm modes only. Upon substitution of the Bogoliubov decomposition into the Eqs.~\eqref{Eq:G2_n} and \eqref{Eq:G2_s}, the density correlation functions can be written to the leading order in the fluctuations as:
\begin{align}
	&{G_r^{(2)}(x,x')} =\left[\Psi_0^*(x)\Psi_0^*(x') \left<\delta\hat{\Psi}_r(x')\delta\hat{\Psi}_r(x)\right>	\right.\nonumber\\
		&+\left.\Psi_0^*(x)\Psi_0(x') \left<\delta\hat{\Psi}_r^\dag(x')\delta\hat{\Psi}_r(x)\right>	+ \text{c.c.}\right].
\label{Eq:G2_1}
\end{align}
The order parameter of the system evolves in time according to the Gross-Pitaevskii equation (GPE):
\begin{equation}
i\hbar\frac{\partial\Psi_0}{\partial t} = \hat{H}_{GP} \Psi_0,
\label{Eq:GPE}
\end{equation}
where $H_{\rm GP}=-\hbar^2\partial_x^2/2m+(g_d/2) |\Psi_0(x,t)|^2+V(x,t)$ is the Gross-Pitaevskii Hamiltonian. The evolution of the Bogoliubov modes is governed instead by the Bogoliubov-de Gennes equations \cite{Castin-BogNumCons-1998}
\begin{equation}
	i\hbar\frac{d}{dt}
	\begin{pmatrix}
    u_r\\
	v_r
  \end{pmatrix}=\mathcal{L}_r
  	\begin{pmatrix}
    u_r\\
	v_r
  \end{pmatrix} 
  =
  \begin{pmatrix}
    L_{QQ}^r     & L_{QQ^*}^r\\
    L_{Q^*Q}^r      & -L_{QQ}^r
  \end{pmatrix}
    	\begin{pmatrix}
    u_r\\
	v_r
  \end{pmatrix},
  \label{Eq:BogOperator}
\end{equation}
in which the (operator-valued) components of the Bogoliubov operator $\mathcal{L}_r$ are defined as
\begin{subequations}
\begin{align}
	L_{QQ}^r&=\left[ H_{\rm GP}+\frac{g_r}{2}\, Q|\Psi_0(x,t)|^2 Q-\mu\right],\label{Eq:LQQ}\\
	L_{QQ^*}^r&=\frac{g_r}{2}\, Q\Psi_0^2(x,t) Q^*,\label{Eq:LQQ*}\\
	L_{Q^*Q}^r&=\left(L_{QQ^*}\right)^*.\label{Eq:LQ*Q}
\end{align}
\label{Eq:LComponents}
\end{subequations}
Here the operator $Q\equiv \mathds{1}-\ket{\Phi_0}\bra{\Phi_0}$ (with $\mathds{1}$ the identity operator) is the projector onto the non-condensed component, that is onto the Hilbert sub-space spanned by all single particle states orthogonal to the condensate wavefunction, and $\mu = -(\hbar^2/2m)(\partial_{x}^2\Psi_0/\Psi_0)+V(x,t=0)+(g_d/2)|\Psi_0|^2$ is the chemical potential of the system calculated for the system at equilibrium. 

\begin{figure*}[!htbp]
\centering
{\subfigure
{\includegraphics[width=0.32\textwidth]{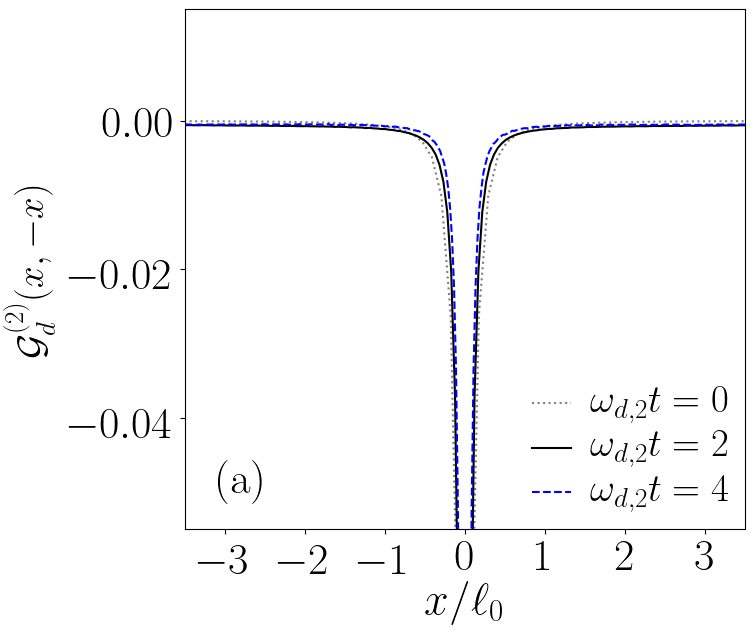}}}
{\subfigure
{\includegraphics[width=0.32\textwidth]{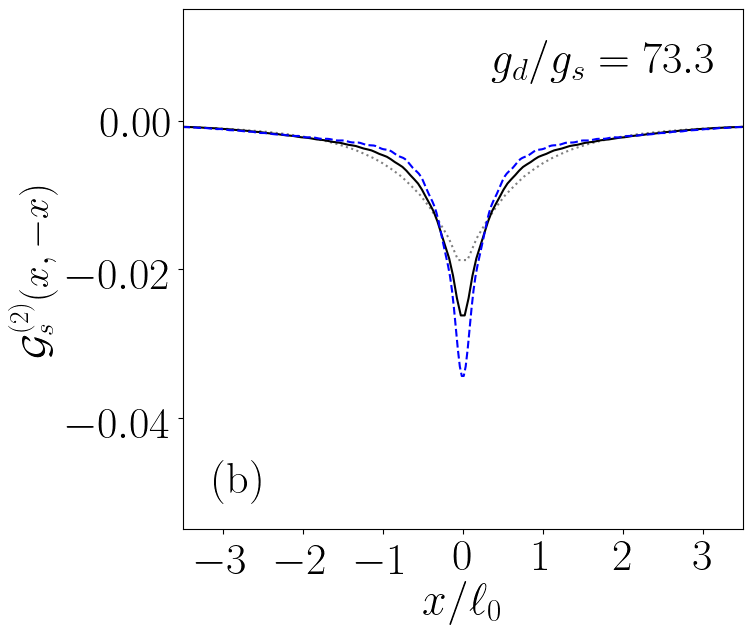}}}
{\subfigure
{\includegraphics[width=0.32\textwidth]{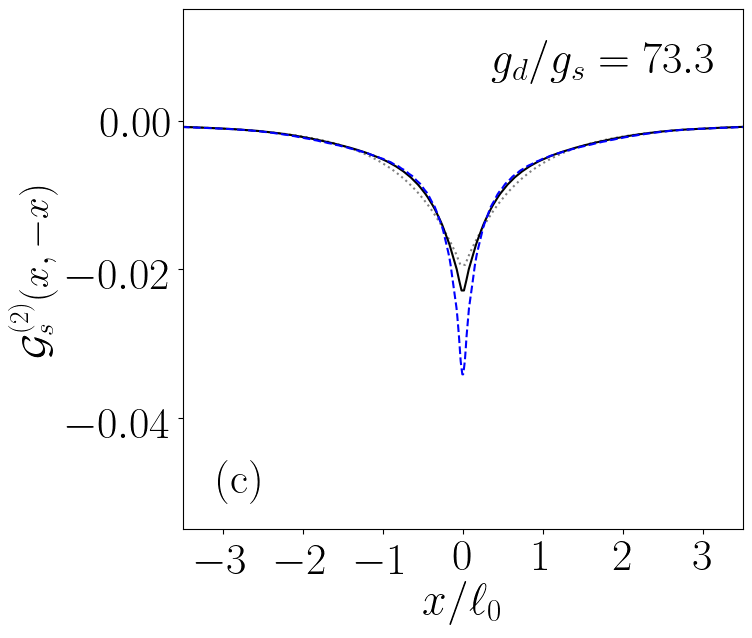}}}
%{\subfigure
%{\includegraphics[width=0.3\textwidth]{bbExp.png}}}
%{\subfigure
%{\includegraphics[width=0.3\textwidth]{bbOsc.png}}}
\caption{Time evolution of the density-density correlation function for a one-dimensional, harmonically trapped two-component condensate of chemical potential $\mu/\omega_0 = 28.25$. Panel (a) shows the contribution to the correlations due to the density excitations, for the case of a linearly expanding system. In panels (b) and (c) we report the contribution to the correlations due to the spin excitations, respectively for the linearly and exponentially expanding systems.}
\label{Fig:5}
\end{figure*}

We carry out our investigation by considering a 1D system whose chemical potential is equal to $\mu/\omega_0 = 28.25$ and by using two different values for the effective spin and density interaction strengths such that $g_d/g_s = 13.3$ and $g_d/g_s = 73.3$. We calculate the density and spin spectra relative to these configurations by numerically diagonalizing the corresponding Bogoliubov operators in Eq.~\eqref{Eq:BogOperator}. These are reported in Fig.~\ref{Fig:3}. We notice that, in the two chosen configurations, the breathing density mode of frequency $\omega_{d,2}$ is close to resonance with the spin modes of angular frequencies $\omega_{s,3}^{13.3}$ and $\omega_{s,5}^{73.3}$, with a detuning from resonance approximately equal to $\pm 7 \%$  and $\pm 2 \%$, respectively. In the notation we use, we indicate in the subscript the quantum number of the modes, while in the superscript the values of the ratio $g_d/g_s$. We expect to see the signature of such resonances in the time evolution of two-body correlations for the oscillating system. 

 We evolve the two-body correlations in time by solving for the time evolution of the background condensate $\Psi_0$ and for the Bogoliubov excitations modes $\{u_r, v_r\}$, by using Eqs.~\eqref{Eq:GPE} and \eqref{Eq:BogOperator}, respectively. Given these solutions we construct the density-density correlations at each time, according to Eqs.~\eqref{Eq:G2_1}.
As in the previous section, the oscillating condensate is implemented by modulating the frequency of the trapping potential in time as $\omega(t)/\omega_0=1+A\exp(-(t-t_0)^2/(2\sigma^2))$. The linearly and exponentially expanding configurations are instead implemented by simply switching off and reverting the trapping potential, respectively. 

We show in Fig.~\ref{Fig:5}(a-c) the results we obtained for the correlations in the expanding case. In order to single out the non-trivial dynamics on top of to the overall expansion, we report the scaled quantity
\begin{equation}
\mathcal{G}^{(2)}_r(y_1,y_2) \equiv \mathcal{V}^2(t)G^{(2)}_r(y_1/\ell(t),y_2/\ell(t)).
\end{equation}
We find that, in the case of an expanding system, the correlations appear featureless. The only noticeable feature is a slight variation in size of the width and the depth of the anti-bunching stripe.

\begin{figure*}[!htbp]
\centering
{\subfigure
{\includegraphics[width=0.32\textwidth]{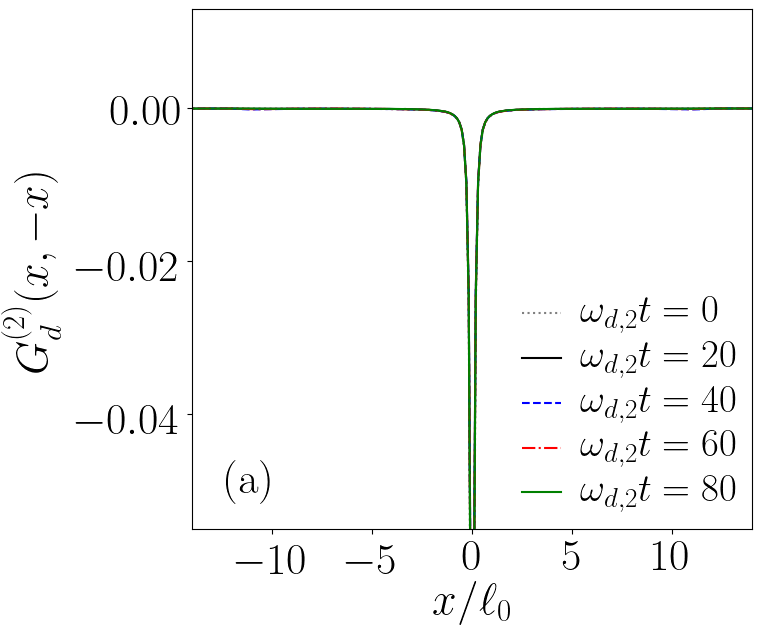}}}
{\subfigure
{\includegraphics[width=0.32\textwidth]{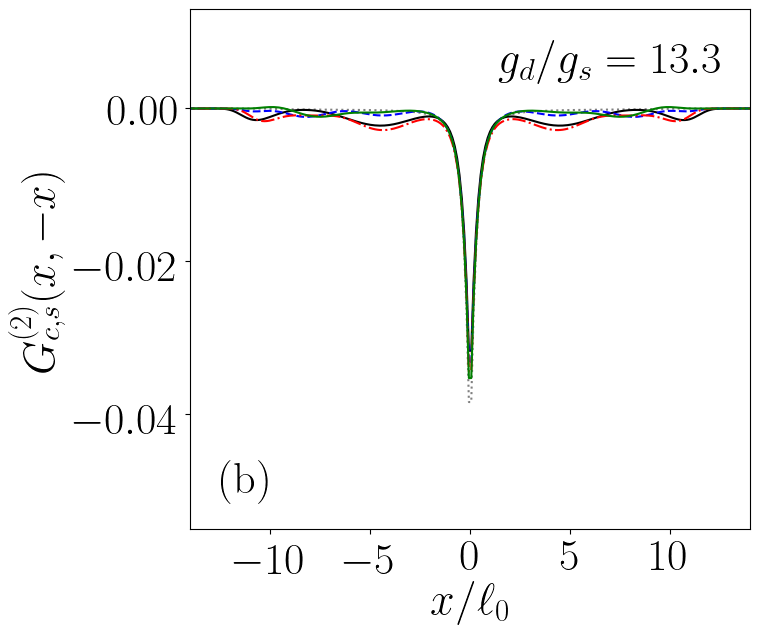}}}
{\subfigure
{\includegraphics[width=0.32\textwidth]{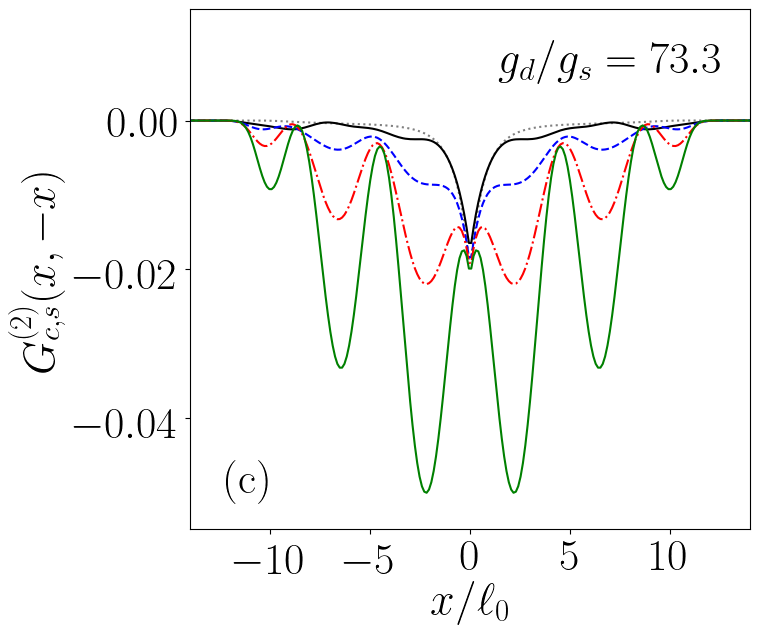}}}
\caption{Time evolution of the density-density correlation function for a one-dimensional, harmonically trapped two-component condensate, of chemical potential $\mu/\omega_0 = 28.25$ whose overall density is oscillating in the breathing mode. In panel (a) we report the  contribution to the correlations due to the density excitations. We see that the zero-point fluctuations that populate these modes are not excited and the correlations do not evolve in time. In panels (b,c) is reported the contribution  to the correlations due to the spin excitations. We clearly see here the structure of the resonant mode, whose vacuum fluctuations are parametrically amplified because of the oscillations in the density.}
\label{Fig:4}
\end{figure*}

The results in Fig.~\ref{Fig:4}(a-c) show instead the much richer dynamics of the spin correlations in the case of the oscillating system. On one hand, in Fig.~\ref{Fig:4}(a) we see that $G_{d}^{(2)}(x,x')$ does not evolve, as expected, because there are no density modes that can be resonantly amplified by density oscillations of the system in the breathing mode. On the other hand, in Figs.~\ref{Fig:4}(b,c) we present the time evolution of spin correlations $G_{s}^{(2)}(x,x')$ for two parameter choices differing for the distance from resonance. In both cases, the zero-point fluctuations in the spin modes are parametrically amplified by the density oscillations. While the effect is relatively weak in the off-resonance case of panel (b) for $g_+/g_-=13.3$, a dramatic resonant enhancement is visible in panel (c) for $g_d/g_s = 73.3$. As time proceeds, the parametric excitation of the resonantly selected spin mode is visible in a monotonically growing amplitude of the spatially oscillating pattern in the spin density correlation function, whose shape is determined by the resonantly selected mode. Other choices of the interaction constant ratio $g_d/g_s$ and of the excited density mode may be used to resonantly address other spin modes, which would result in a different spatial pattern of the spin correlation function.

\section{Conclusions\label{Sec:Conclusions}}

In this work we have theoretically studied the analog of cosmological particle creation in a non-stationary Universe, using an analog model based on a two-component Bose-Einstein condensate. We have shown that the collective spin excitations of the system behave as a quantum field experiencing a time-dependent background determined by the time-dependent density profile. As a result of this time modulation, the zero-point vacuum fluctuations in the spin modes can be parametrically amplified according to a quantum particle creation process. 
By working in the Thomas-Fermi limit, we developed a theoretical model that is able to analytically describe the dynamics of the quantized collective excitations on top of classical mean-field condensate. Our theoretical predictions have been then validated by a full ab initio numerical study of the time evolution of the quantum fluctuations in a inhomogeneous condensate. 

In the perspective of the experimental investigation of this physics, we have focused our attention on the spatial correlation function of spin-fluctuation. On one hand, no specific  feature witnessing the particle creation appears in the case of an expanding condensate, mostly due to an effective friction analogous to the Hubble friction in Cosmology. On the other hand, unambiguous signatures are visible in the case of an oscillating condensate. Because of the onset of a resonant interaction between the density oscillations and certain spin modes, the density correlations develops a peculiar oscillating pattern that is very promising in view of experimental observations with state-of-the-art cold atom technology. 

A direct next step of our work will be to extend our study in the presence of a coherent coupling between the two species, so to provide an effective mass to the spin modes~\cite{Visser2005,Abad2013,Butera2017} and investigate particle creation effects for massive fields~\cite{Visser2005}.
On a longer term, our proposal opens exciting perspectives in the direction of studying back-reaction phenomena.
In contrast to most previous works, here this amplification is not induced by externally modulating in time a physical parameter of the system, but rather originates from the dynamical evolution of the system itself. As a result, the background is no longer externally imposed as in traditional quantum field theories on curved space-times~\cite{birrell1984quantum}, but is a fully-fledged degree of freedom of the problem. This feature holds a great promise in view of studying how the parametrically excited quantum field back-reacts onto the background and modifies its dynamics, e.g. by inducing a friction onto the density oscillations~\cite{Robertson-PRD-PreHeatAn-2019,Butera-BR_DCE-2019}. Understanding such back-reaction phenomena in condensed-matter toy models provides a promising avenue to shine light on a number of questions of cosmological interest, related for example to the early inflationary stage of the Universe, or the ultimate stage of existence of a black hole. 

\section{Acknowledgements}
Continuous stimulating discussions with Gabriele Ferrari, Alessio Recati, Anna Berti and Luca Giacomelli. are warmly acknowledged.
S.~B. acknowledges funding from the Leverhulme Trust Grant No. ECF-2019-461 and the Lord Kelvin/Adam Smith (LKAS) Leadership Fellowship. I.C. acknowledges support from the European Union Horizon 2020 research and innovation program under Grant Agreement No. 820392 (PhoQuS) and from the Provincia Autonoma di Trento.

\bibliography{QFTCS-BR.bib}

\end{document}